\documentclass[a4paper,11pt]{article}
\pdfoutput=1 

\usepackage{jcappub} 

\usepackage[T1]{fontenc} 
\usepackage{xcolor}
\definecolor{lightsabergreen}{rgb}{.14,.64,.14}
\definecolor{lightgreen}{rgb}{.14,.44,.14}

\title{\boldmath The Effect of Multiple Cooling Channels on the Formation of Dark Compact Objects }

\author[a,b,c]{Joseph Bramante,}
\author[a,b,c]{Melissa Diamond,}
\author[a,b]{and J. Leo Kim}

\affiliation[a]{Department of Physics, Engineering Physics \& Astronomy, Queen's University, \\ 64 Bader Lane, Kingston, ON K7L 3N6, Canada}
\affiliation[b]{Arthur B. McDonald Canadian Astroparticle Physics Research Institute,\\ 64 Bader Lane, Kingston, ON K7L 3N6, Canada}
\affiliation[c]{Perimeter Institute for Theoretical Physics,\\
31 Caroline St N, Waterloo, ON N2L 2Y5, Canada}

\emailAdd{joseph.bramante@queensu.ca}
\emailAdd{m.diamond@queensu.ca}
\emailAdd{leo.kim@queensu.ca}

\abstract{A dissipative dark sector can result in the formation of compact objects with masses comparable to stars and planets. In this work, we investigate the formation of such compact objects from a subdominant inelastic dark matter model, and study the resulting distributions of these objects. In particular, we consider cooling from dark Bremsstrahlung and a rapid decay process that occurs after inelastic upscattering. Inelastic transitions introduce an additional radiative processes which can impact the formation of compact objects via multiple cooling channels. We find that having multiple cooling processes changes the mass and abundance of compact objects formed, as compared to a scenario with only one cooling channel. The resulting distribution of these astrophysical compact objects and their properties can be used to further constrain and differentiate between dark sectors.}

\begin{document}
\maketitle
\flushbottom

\section{Introduction}
\label{sec:intro}

Despite the overwhelming evidence for dark matter (DM) provided by its gravitational interactions, its true nature remains unknown. Cold dark matter (CDM) has been the leading DM model for decades due to its success in describing the large scale structure of the Universe. 
However, on smaller galactic scales there is an evolving interplay between CDM predictions and observation. This interplay first appeared in cored galaxies that CDM predicted would be cuspy \cite{Moore:1994yx, Flores:1994gz} and populations of apparently missing satellite galaxies \cite{Klypin:1999uc, Moore:1999nt}. As astronomers begin to probe the DM distribution on smaller scales \cite{Oh:2015xoa,Hezaveh:2016ltk,Obuljen:2020ypy}, the success of the early simple CDM models has come under increasing scrutiny, and now it is common to also study baryonic feedback and related effects that could account for the departure from CDM models \cite{coretocusp,baryonicfeedbacksmallscale} and revisit searches for satellite galaxies \cite{nomissingsatellites}.  These astrophysical observations have also inspired the search for DM models which leave the large-scale successes of CDM intact while modifying small-scale structure in the dark sector (see Ref. \cite{Tulin:2017ara} for a comprehensive review of self-interacting dark matter, as one example of a non-CDM framework). Here we will study a self-interacting dark sector that is dissipative and able to form compact objects.

A dark sector able to form dark compact objects would fundamentally alter our understanding of the distribution of small-scale astrophysical objects. Several different types of dark structures and formation mechanisms have already been proposed; for example, primordial black holes  \cite{Hawking:1974rv, Carr:1974nx}, axion stars \cite{Tkachev:1991ka, Eby:2016cnq},   and dissipative dark sectors \cite{Fan:2013yva, Fan:2013tia, McCullough:2013jma, Fan:2013bea, Foot:2016wvj, Buckley:2017ttd, Chang:2018bgx, Huo:2019yhk, Dvali:2019ewm, Shen:2021frv, Shen:2022opd, Ryan:2022hku, Gurian:2022nbx, Roy:2023zar, Gemmell:2023trd}. The last case arises from the natural possibility that DM is not just one particle, but rather a sector comprised of a variety of dark particles, analogous to the photons, electrons, and baryons of the Standard Model. Depending on the structure of the dark sector, one may imbue it with several novel interactions. Radiative processes in dark sectors in particular, have exciting implications. For instance, radiative cooling in gasses of dark particles could influence the formation and evolution of structure throughout their cosmological history. This is analogous to the formation of compact objects in the baryonic sector, which is precisely due to cooling mechanisms which result in the loss of kinetic energies for baryons -- allowing them to descend further into their gravitational wells and eventually form stars and planets. If compact dark objects exist, then they may similarly be formed through dissipative processes.

Often dissipative dark sectors are sufficiently complex that their behavior must be modeled numerically. For instance, atomic dark matter \cite{Kaplan:2009de}, in which the dark sector is comprised of a dark proton, a dark electron, and a dark photon, is a model that permits both atomic and molecular processes -- resulting in a plethora of relevant interactions one must account for when considering the formation histories of  the associated astrophysical structures. Recent studies have investigated the complicated structure formation from atomic dark matter by either studying the range of possible simulation results by interpolating between a range of assumptions regarding density evolution \cite{Ryan:2021dis, Gurian:2021qhk, Ryan:2021tgw, Gurian:2022nbx} or using numerical hydrodynamics simulation suites \cite{Springel:2005mi,Hopkins:2014qka} to study these objects \cite{Roy:2023zar}.

In this work, rather than considering a complex dark sector with a great many of processes, we aim to minimally elaborate on dark matter cooling in a simple thermodynamic framework as in, $e.g.$ \cite{Buckley:2017ttd,Chang:2018bgx}. In particular, we will follow the treatment of Ref. \cite{Chang:2018bgx}, which studied the formation and evolution of exotic compact objects using a set of semi-analytic processes, which did not require numerical simulations. While their analysis was performed considering a simple dark sector with only one relevant cooling channel, we investigate the implications of having multiple cooling channels in a similar, semi-analytic treatment. For concreteness, we will work with an asymmetric, subdominant, inelastic dark sector comprised of a dark fermion (with two mass states) and dark photon, and show how the additional cooling channel provided by the inelastic transition between fermion states leads to a markedly different evolution of dark sector dynamics in forming dark compact objects. Our results build on prior work, and we find it is possible to obtain a somewhat wider distribution of dark compact objects ranging from pressure-supported, asymmetric dark stars \cite{Kouvaris:2015rea, Eby:2015hsq} to dark black holes that can span several orders of magnitude in mass. These dark black holes are especially of interest, since we find the inelastic dark sector (with a lower Chandresekhar mass than in the Standard Model) can form sub-solar mass black holes \cite{Shandera:2018xkn}, which are not expected from any known non-primordial astrophysical processes in the baryonic sector\footnote{While primordial black holes are able to have masses below a solar mass and would not necessarily invoke the existence of a new particle beyond the standard model, their existence would invoke non-conventional physics around the epoch prior to Big Bang Nucleosynthesis. }. The existence of such low mass black holes could serve as smoking gun signatures of physics beyond the standard model if detected \cite{Bramante:2017ulk,Dasgupta:2020mqg}.

The outline of this paper is as follows. In Section \ref{sec:cooling} we will describe a simple dark sector and its radiative processes, and obtain the relevant cooling rates for the subdominant dissipative dark matter halo. In Section \ref{sec:formation}, we will then use the cooling rates obtained in Section \ref{sec:cooling} along with a semi-analytical formalism which arises from thermodynamical arguments to understand the formation and evolution of these dark compact objects. Then, in Section \ref{sec:landscape} we will show what a possible distribution of dark compact objects could look like in the presence of multiple cooling channels, and discuss the implications and prospects of these objects as observables. Finally, we will summarize our results and make concluding remarks in Section \ref{sec:conclusion}.

\section{Multiple Cooling Channels} \label{sec:cooling} 

In the next section, we will investigate the formation of dark exotic compact objects by studying the evolution of a gas comprised of inelastic dark matter particles and dark photons. To do so, we must first discuss the relevant processes that our particles undergo in a dark sector gas cloud. We are particularly interested in the non-equilibrated ($i.e.$ inelastic) dissipative processes that can transfer energy out of the gas cloud, lower the kinetic energy of the gas particles and thus cool the overall gas. Since we are interested in how compact object formation differs in models with multiple cooling channels from those with one, we chose a model with two cooling mechanisms.

There are several mechanisms that produce interesting dark sectors with multiple relevant cooling channels for dark structure formation. In what follows, we select a specific dissipative dark sector for concreteness. We consider a simple, asymmetric, dark sector consisting of a dark Dirac fermion $\chi$ and a massive gauge photon $A^\mu$. In what follows, we use the notation $\gamma_D = \gamma_\mu A^\mu$ and refer to $\gamma_D$ as the dark photon for the sake of brevity. We assume the dark fermion has a mass splitting, resulting in a ground state mass $\chi_1$ and an excited state mass $\chi_2$ for the dark fermion, with mass difference $\delta = m_{\chi_2} - m_{\chi_1}$. In counter-distinction to some inelastic dark matter models \cite{Tucker-Smith:2001myb,Tucker-Smith:2004mxa,Batell:2009vb,Bramante:2016rdh,Bramante:2020zos}, we also endow our theory with interactions that permit the dark photon to interact elastically with $\chi_1$ and $\chi_2$, which implies Lagrangian parameters at tree-level not only of the form $\mathcal{L} \supset g_D \chi_1 \gamma_D \chi_2 $, but also of the form $\mathcal{L} \supset g_D \chi_1 \gamma_D \chi_1 $ and $\mathcal{L} \supset g_D \chi_2 \gamma_D \chi_2 $, so that $\chi_1$ and $\chi_2$ will have a tree-level Bremmstrahlung interaction with other $\chi$ particles, and where $g_D$ is the gauge coupling constant corresponding to fine structure constant $\alpha_D \equiv \frac{g_D^2}{4 \pi}$. Note that if we followed the rubric of some past dark photon-mediated inelastic dark matter models ($e.g.$ \cite{Batell:2009vb,Bramante:2016rdh,Bramante:2020zos}), we would not obtain vertices with $\chi_1 \gamma_D \chi_1 $ interactions at tree-level since the dark photon gauge interaction terms in the Lagrangian would be off-diagonal with respect to the fermion states $\chi_1,\chi_2$. Rather, we expect these kinds of interactions to occur more prominently in models of composite inelastic dark matter \cite{Alves:2009nf,SpierMoreiraAlves:2010err,Kribs:2016cew,Bramante:2016rdh,Cline:2021itd}, where the mass splitting is generated either by the non-perturbative spectrum of composite states or mass differences between composite constituents.  Indeed, another example of a theory with both diagonal and off-diagonal interactions between $\chi_1,\chi_2$ is the spectrum of excited baryonic states in the Standard Model: the delta baryon $\Delta^+$ in the Standard Model, which as the spin-excited state of the proton, has both elastic and inelastic interactions with the Standard model photon in the baryonic sector. There are also non-composite models where the mass splitting is generated by additional gauge groups, and so diagonal and off-diagonal couplings between $\chi_1,\chi_2$ are generated, see $e.g.$ \cite{Ko:2022kvl}. 

Before continuing, we emphasize that the choice of model here is important only in that it provides both an inelastic and elastic long-range scattering process, thereby creating several distinct dissipative mechanisms. We reiterate that we have only selected a specific model for the purpose of concreteness, and that the many results of this work can be applied to any dissipative dark sector with multiple cooling channels. The inclusion of two distinct mass states permits us to study inelastic dissipative processes in a simple framework. This model could be realized in any number of dark matter models frameworks in which tree-level inelastic processes are relevant, which have been extensively studied in \cite{Tucker-Smith:2001myb, Tucker-Smith:2004mxa, Finkbeiner:2007kk, Arkani-Hamed:2008hhe, Batell:2009vb,Alves:2009nf, Bramante:2016rdh, Bramante:2020zos,Ko:2022kvl}. In this work, we will work with a mass hierarchy where $m_{\gamma_D} \ll \delta \ll m_{\chi_1}$. Since the mass splitting $\delta \ll m_{\chi_1}$ in our setup, we can sometimes approximate the two mass states as being roughly equal in what follows, and in this case we will write $m_\chi \simeq m_{\chi_1}$ for the sake of brevity sometimes when we are referring to (ground state) inelastic dark matter particles. 

To continue, we will be interested in the upscattering processes $\chi_1~ \chi_1 \to \chi_1~ \chi_2$ and $\chi_1~\chi_1 \to \chi_2 ~\chi_2$. For each process, we will refer to any associated quantities with subscripts IE,1 and IE,2 respectively. After the initial upscattering process, so long as the decay times for $\chi_2 \to \chi_1 ~ \gamma_D$ are very quick, the resulting $\chi_2$ particles decay swiftly back into $\chi_1$ particles and emit a dark photon with energy $\sim \delta$, thereby cooling the inelastic dark matter halo. In what follows, we assume this to be the case, and that the decay times are much smaller than the upscattering times, such that the upscattering times are the limiting timescales for cooling. We note that if the decay times for $\chi_2 \to \chi_1 ~ \gamma_D$ were sufficiently comparable to the upscattering times, we could end up with a population of $\chi_2$ particles in the inelastic dark matter halo, resulting in a more complicated cooling process. We leave this case to future work. The cross sections for these upscattering processes are given by (see $e.g.$ \cite{Schutz:2014nka,Bramante:2016rdh})
\begin{align}
    \sigma_{\mathrm{IE},1} &= \frac{ 4 \pi \alpha_D^2 m_\chi^2 \sqrt{1 - \frac{\delta}{ m_\chi v^2 } } }{ m_{\gamma_D}^4 \left[ \left( 1 - \delta m_\chi/m_{\gamma_D}^2 \right)^2 + 4 m_\chi^2 v^2/ m_{\gamma_D}^2 \right] }, \\
    \sigma_{\mathrm{IE},2} &= \frac{ 4 \pi \alpha_D^2 m_\chi^2 \sqrt{1 - \frac{2 \delta}{ m_\chi v^2 } } }{ m_{\gamma_D}^4 \left[ \left( 1 - 2 \delta m_\chi/m_{\gamma_D}^2 \right)^2 + 4 m_\chi^2 v^2/ m_{\gamma_D}^2 \right] },
\end{align}
where $v \simeq \sqrt{3 T_\chi/m_\chi }$ is an estimate for the velocity of the inelastic dark matter particles inside of the halo given the temperature of the inelastic dark matter gas, $T_\chi$. Note that while the two interactions vertices $\chi_1 \gamma_D \chi_1 $ and $\chi_1 \gamma_D \chi_2 $ could have different couplings, we assume that they have the same coupling strength $\alpha_D$ for simplicity.

Given these cross sections, we can compute two inelastic cooling rates for our inelastic dark fermions colliding in a dark gas cloud,
\begin{align}
    \Gamma_{\mathrm{IE},1} = \delta n_{\chi} \sigma_{\mathrm{IE},1} v ,  \qquad 
    \Gamma_{\mathrm{IE},2} = 2 \delta n_{\chi} \sigma_{\mathrm{IE},2} v,
    \label{eq:Gamma_IE}
\end{align}
where $n_\chi$ is the number density of the inelastic dark matter particles in our inelastic dark matter halo. The factors $\delta$ and $2\delta$ for $\Gamma_{\mathrm{IE},1}$ and $\Gamma_{\mathrm{IE},2}$ respectively are the total energies carried away by the dark photons after the $\chi_2$ particles decay into $\chi_1$ particles. We note that while the energy carried away by the dark photons is $\sim \delta$ so long as the kinetic energy of the $\chi_2$ particles is negligible, this is not necessarily true for $T \gtrsim m_\chi$, $i.e.$ in the relativistic regime. However, since we restrict our analysis to a regime in which the kinetic energy of $\chi_2$ is non-relativistic regime, we can assume that the dark photons produced in $\chi_2$ decay carry energy $\delta$ with negligible corrections.

Hence the overall cooling rate is given by a combination of all of the cooling processes, which can be written as
\begin{align}
    \Gamma_\mathrm{cool} = (\Gamma_\text{Compton} + \Gamma_\text{BS}  ) e^{-V^{1/3} \sqrt{N_{\mathrm{sc}}} / \ell_{\gamma_D}^{\mathrm{abs, BS}} } + (\Gamma_\mathrm{IE,1} + \Gamma_\mathrm{IE,2}) e^{-V^{1/3} \sqrt{N_{\mathrm{sc}}} / \ell_{\gamma_D}^{\mathrm{abs, IE}} } , 
    \label{eq:Gamma_new}
\end{align}
where the exponential factor is to account for dark photon rescatterings, with the number of rescatterings given by
\begin{align}
    N_{sc} = \left( \frac{V^{1/3}}{\ell_{\gamma_D}^C} \right)^2.
\end{align}
The number of rescatterings is computed using the volume of the inelastic dark matter halo $V = M/\rho_\chi$, where $M$ is the mass of the halo, and the mean free path for Compton scattering,
\begin{align}
    \ell_{\gamma_D}^C = \frac{3 m_\chi^2}{8 \zeta \pi n_\chi \alpha_D^2  },
\end{align}
where $\zeta = 1$ or $2/3$ for polarization-averaged cross sections of relativistic or non-relativistic dark photons, respectively. The two quantities $\ell_{\gamma_D}^\mathrm{abs, BS}$ and    $\ell_{\gamma_D}^\mathrm{abs, IE}$ that appear in Eq. \eqref{eq:Gamma_new} are the absorption mean free paths for Bremsstrahlung-produced dark photons and inelastically produced dark photons respectively, given by
\begin{align}
    \ell_{\gamma_D}^\mathrm{abs, BS} = G(T_\chi, T_{\chi}) \frac{(m_{\chi} T_{\chi})^{5/2} }{ n_{\chi}^2 \alpha_D^3 }, \label{eq:mfp_BS}
\end{align}
for the Bremsstrahlung-produced dark photons, and
\begin{align}
    \ell_{\gamma_D}^\mathrm{abs, IE} = G( \delta, T_{\chi}) \frac{(m_{\chi} T_{\chi})^{5/2} }{ n_{\chi}^2 \alpha_D^3 }, \label{eq:mfp_IE}
\end{align}
for the inelastically produced dark photons. In the inelastic expression, we have indicated the expected energy for the dark photons produced through de-excitation of the inelastic state, $\omega_{D} \approx \delta$, where this is appropriate in the limit $m_{\gamma_D} \ll \delta$. The function $G(x)$ that appears in both of the mean free paths is given by \cite{Haug:1975}
\begin{align}
    G(\omega_D, T_{\chi}) = \left[ \frac{16 \pi^{3/2}}{15} \frac{T_{\chi}}{\omega_D} e^{\omega_D/T_{\chi}} \int_0^1 dx ~ \frac{F(x)}{x^3} e^{-\omega_D/(T_{\chi}x)} \right]^{-1}, \label{eq:G_x}
\end{align}
where $F(x)$ is given as
\begin{align}
    F(x) = \left[ 17 - \frac{3x^2}{(2-x)^2} \right] \sqrt{1-x} + \frac{12(2-x)^4 - 7x^2(2-x)^2 - 3x^4 }{(2-x)^3} \ln \left[ \frac{1+  \sqrt{1-x}}{\sqrt{x}} \right].
\end{align}
With these definitions, while Eq. \eqref{eq:mfp_IE} must be numerically solved, Eq. \eqref{eq:mfp_BS} simplifies and can be analytically solved to yield
\begin{align}
    \ell_{\gamma_D}^\mathrm{abs, BS} = 3.0 \times 10^{-3} \frac{(m_{e_D} T_{e_D})^{5/2} }{ n_{e_D}^2 \alpha_D^3 },
\end{align}
which is slightly different from the result in Ref. \cite{Chang:2018bgx}, which we believe arises due to a typo in simplifying their expression for Eq. \eqref{eq:G_x}. For more details on the derivation of the absorption mean free paths, we point interested readers to Appendix B in Ref. \cite{Chang:2018bgx} for further discussion.

The non-relativistic Bremsstrahlung cooling rate is given in Ref. \cite{Chang:2018bgx} as
\begin{align}
    \Gamma_\text{BS} = \frac{32 \alpha_D^3 \rho_{\chi} T_{\chi} }{ \sqrt{\pi} m_{\chi}^3 } \sqrt{\frac{T_{\chi}}{ m_{\chi} }} e^{-m_{\gamma_D}/T_{\chi}}, \label{eq:Gamma_BS}
\end{align}
where $\rho_\chi$ is the energy density of the inelastic dark matter halo. Note that technically, Eq. \eqref{eq:Gamma_new} is dependent on the cooling rate $\Gamma_\text{Compton}$ for Compton scattering of the inelastic dark matter particles with dark CMB photons. However, similar to Ref. \cite{Chang:2018bgx}, we will work in the late Universe where $ \Gamma_\text{Compton} \ll \Gamma_{\text{BS}}$, and thus we can effectively ignore the contributions from 
$\Gamma_\mathrm{Compton}$. 

\begin{figure}[tbp]
\centering 
\includegraphics[width=.49\textwidth]{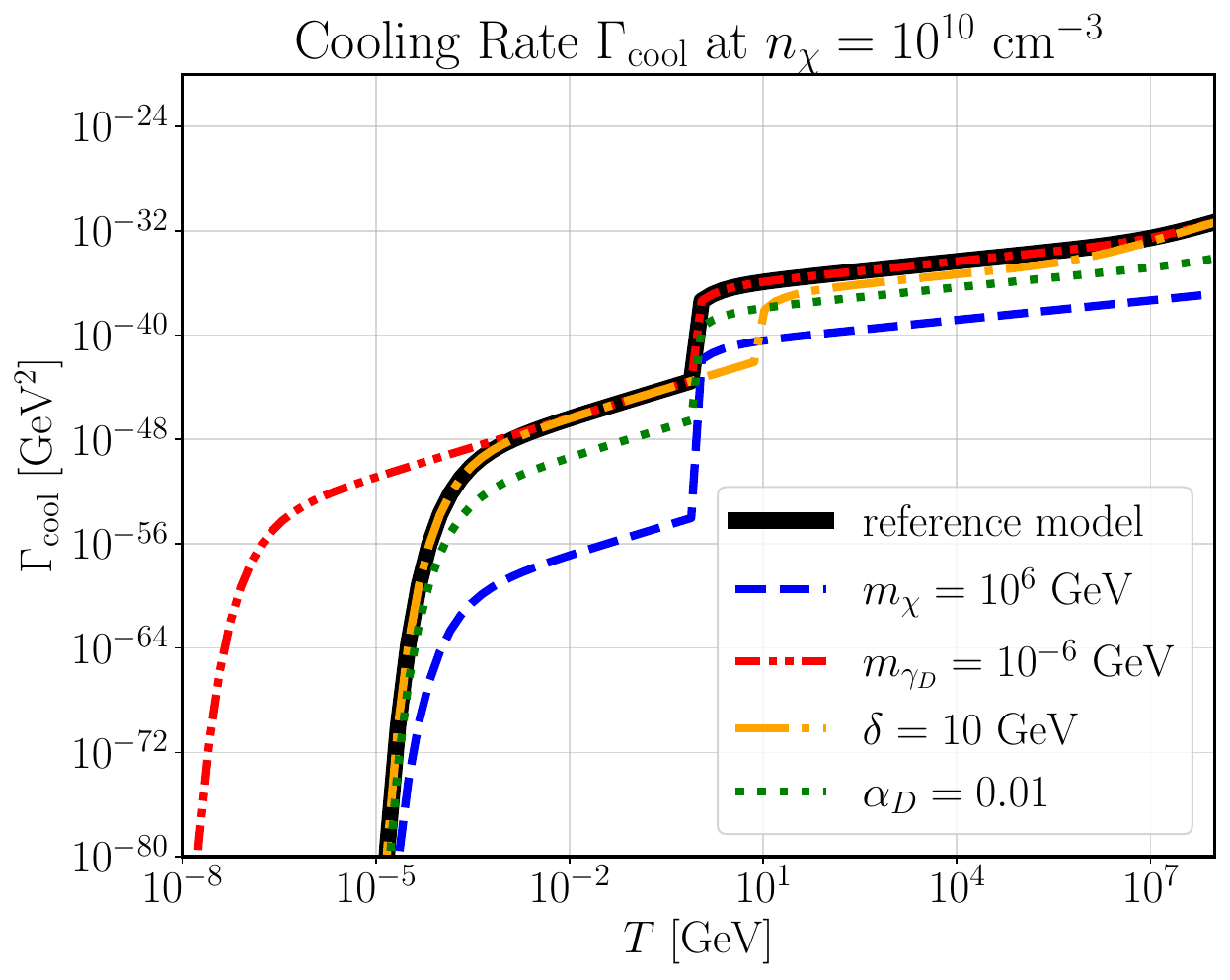}
\includegraphics[width=.49\textwidth]{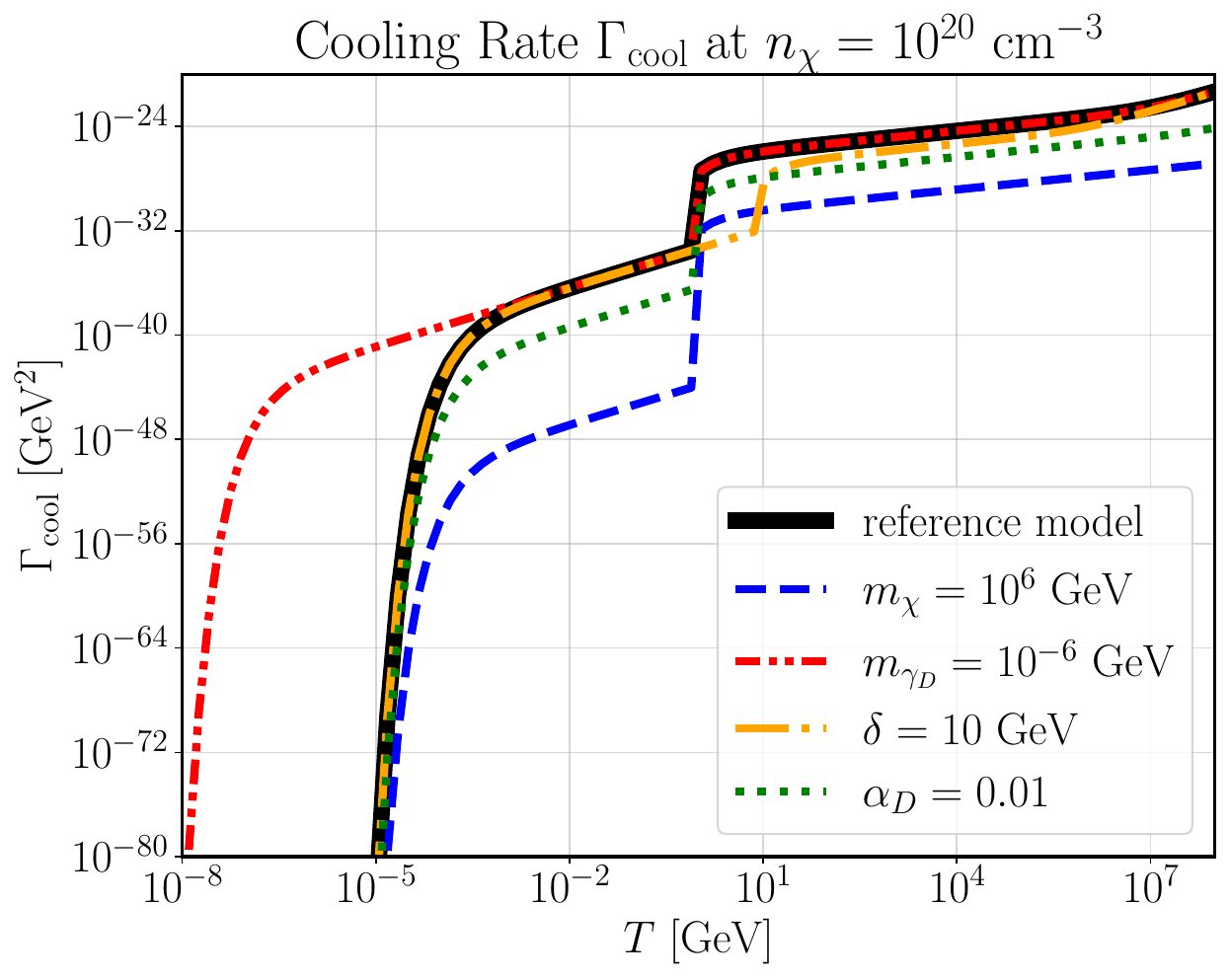}
\caption{Overall cooling rate $\Gamma_\mathrm{cool}$ given in Eq. \eqref{eq:Gamma_new} as a function of temperature $T_\chi$ at a fixed number densities $n_{\chi} = 10^{10}$ cm$^{-3}$ in the left panel and $n_\chi = 10^{20}$ cm$^{-3}$ in the right panel. The black solid curve is the so-called reference model, with model parameters $m_\chi = 10^3$ GeV, $m_{\gamma_D} = 10^{-3}$ GeV, $\alpha_D = 0.1$, and $\delta = 1$ GeV. The blue, red, yellow, and green curves represent the cooling rates with just one parameter changed from the reference model: $m_\chi = 10^6$ GeV, $m_{\gamma_D} = 10^{-3}$ GeV, $\delta = 10~{\rm GeV}$, and $\alpha_D = 0.01$ respectively. At low temperatures where $T < \delta$, the dominant cooling process is Bremsstrahlung, while for high temperatures where $T \gtrsim \delta$, the dominant cooling process is from inelastic transitions.} \label{fig:Gamma_fixed_n} 
\end{figure}

One can see the effects of having multiple contributions to the cooling rate $\Gamma_\mathrm{cool}$ in Fig. \ref{fig:Gamma_fixed_n}, which shows the overall cooling rate at a given fixed inelastic dark matter halo number density $n_\chi$ as a function of the gas temperature $T_\chi$. The left panel corresponds to a fixed number density of $n_\chi = 10^{10}$ cm$^{-3}$ and the right panel corresponds to a number density of $n_\chi = 10^{20}$ cm$^{-3}$. Comparing the two panels, the magnitude of the cooling rate increases as the number density increases, which is expected from the linear dependence of the cooling rate on the number density as seen in Eqs. \eqref{eq:Gamma_IE} and \eqref{eq:Gamma_BS}. The cooling rates in the inelastic and Bremsstrahlung channels are suppressed at temperature values lower than $\delta$ and $m_{\gamma_D}$ respectively, due to the presence of the exponential factors in Eqs. \eqref{eq:Gamma_IE} and \eqref{eq:Gamma_BS}. As seen in the figure, the inelastic channel cooling rates are at a much higher amplitude than the Bremsstrahlung rates once the channel is activated. Hence as long as we are in the regime where inelastic cooling is active, it will be the dominant process. In other words, if $T_\chi \gtrsim \delta$, then since $\Gamma_\mathrm{BS} \ll \Gamma_\mathrm{IE}$, the overall cooling rate is dominated by the inelastic cooling channel. On the other hand, for $T_\chi \lesssim \delta$, inelastic cooling is inefficient and has a negligible contribution to the cooling rate due to the exponential suppression in Eq. \eqref{eq:Gamma_new}. Hence in this regime Bremsstrahlung is the only relevant cooling channel.

Before moving onto the next section, we note that while there are two separate processes we end up considering our inelastic cooling model ($i.e.$ $\chi_1 ~ \chi_1 \to \chi_2  ~\chi_2$ and $\chi_1~ \chi_2 \to \chi_2 ~ \chi_2$), the activation energies of each process only differ by a factor of two. As a consequence, we should expect that in a cooling regime dominated by inelastic cooling, the two processes will usually operate as a single effective inelastic cooling channel, since a bath of particles with Boltzmann-distributed temperatures capable of exciting the lower energy transition, will also excite the other, $cf.$ the exponential factor in Equation \ref{eq:Gamma_new}.

\section{Formation of Dark Compact Objects from Inelastic Dark Matter} \label{sec:formation}

\subsection{Details of Collapse}

The formation of the dark compact objects can be divided into a linear regime, in which the initial perturbations grow if necessary conditions are met, and a non-linear regime, where the overdensities can begin to collapse as they decouple from the Hubble flow and become self-gravitating objects. In this we work we have adopted the simplified, semi-analytical treatment for the growth and non-linear collapse of perturbations following Ref. \cite{Chang:2018bgx}, which we briefly review in this subsection.

Turning to the properties of the collapsing objects, we begin by giving a thermodynamic description of the inelastic dark matter gas. The pressure of the inelastic dark matter gas is given by
\begin{align}
    P_{\chi} = n_{\chi} T_{\chi} + \frac{2\pi \alpha_D n_{\chi}^2}{m_{\gamma_D}^2}, \label{eq:pressure}
\end{align}
where the first term corresponds to the kinetic pressure of the inelastic dark matter particles and the second term corresponds to the dark photon repulsive force. The corresponding sound speed $c_s$ for the inelastic dark matter gas can be written as
\begin{align}
    c_s = \sqrt{ \frac{T_\chi}{m_\chi} + \frac{4 \pi \alpha_D n_\chi }{m_\chi m_{\gamma_D}^2} }.
\end{align}
The associated proper Jeans length is given by
\begin{align}
    \lambda_J = c_s \left( \frac{\pi}{\rho G} \right)^{1/2},
\end{align}
where $\rho$ is the density of the perturbation. The growth of perturbations in the linear regime proceeds following the Jeans criterion, in which the initial perturbations grow as long as the proper wavelength of the perturbation $\lambda_P$ is greater than the proper Jeans length, $\lambda_P > \lambda_J$. 

As the perturbations grow in the linear regime, they eventually decouple with the Hubble flow and become non-linear self-gravitating objects. The point at which the perturbations become non-linear is often called turnaround. After turnaround, the evolution of the overdensities continue in a more complicated framework than that of the linear regime.

In the non-linear regime following the growth of perturbations, collapse of the inelastic dark matter perturbation proceeds if the mass of the region is greater than the Jeans mass, given by
\begin{align}
    m_J = \frac{4\pi}{3} \left( \frac{\lambda_J}{2} \right)^3 \rho_{\chi} = \frac{\pi}{6} c_s^3 \left( \frac{\pi}{\rho G} \right)^{3/2}\rho_{\chi}, \label{eq:m_J_full}
\end{align}
where $\rho$ is the total density of everything relevant in both the baryonic sector and the dark sector, and $\rho_\chi$ is the density of the inelastic dark matter halo. 

As evident from Eq. \eqref{eq:m_J_full}, the inelastic dark matter halo Jeans mass is dependent on the total density of both baryonic and dark sectors, and so the evolution of the inelastic dark matter halo is coupled with the evolution of the baryonic sector as well as CDM. However, for simplicity, we will treat the inelastic dark matter halo independently for the following reasons. First, since CDM is the dominant contribution to the overall gravitational potential of the halo, the inelastic dark matter particles and the baryons will collapse towards the center of the halo as they are both able to convert gravitational energy into thermal energy, which is then released through cooling. Once the density of the inelastic dark matter particles and the baryons exceed the density of CDM, they are self-gravitating and CDM can be ignored. Second, although the evolution of the inelastic dark matter halo can be influenced by the presence of baryons, for simplicity we assume that neglecting the baryons does not significantly change our conclusions. We also note that including the gravitational influence of baryons would only help the collapse of the inelastic dark matter halo since it would increase the depth of the gravitational potential for the inelastic dark matter particles to fall into.

Therefore ignoring the influence of baryons and CDM, setting $\rho \simeq \rho_\chi$ in Eq. \ref{eq:m_J_full} allows one to simplify the Jeans mass as
\begin{align}
    m_J = \frac{\pi}{6} c_s^3 \left( \frac{\pi}{G} \right)^{3/2} \left( \frac{1}{\rho_\chi} \right)^{1/2} \label{eq:m_J}.
\end{align}
Since the evolution of the inelastic dark matter halo is assumed to be independent of baryons and CDM, we will only concerned with one differential equation that governs the evolution of the inelastic dark matter halo.

A simple thermodynamical argument results in the following differential equation that relates the evolution of the inelastic dark matter density to its temperature:
\begin{align}
    \frac{d \log T_{\chi} }{ d \log \rho_{\chi} } = \frac{2}{3} \frac{m_{\chi} P_{\chi} }{ \rho_{\chi} T_{\chi} }  - 2 \frac{t_\text{collapse} }{ t_\text{cool} }, \label{eq:DE}
\end{align}
where
\begin{align}
    t_\text{collapse} = \left( \frac{d \log \rho_{\chi} }{dt} \right)^{-1}; \quad t_\text{cool} = \frac{3 T_{\chi}}{\Gamma_\mathrm{cool} }.
\end{align}
The collapse timescale $t_\mathrm{collapse}$ is determined from the current state of collapse of the inelastic dark matter halo from three different stages of collapse, which we summarize qualitatively below (interested readers should also refer to Ref. \cite{Chang:2018bgx}):
\begin{enumerate}
    \item {\bf Adiabatic free-fall}\footnote{Note that adiabatic free-fall is only an approximation of the collapse in this period. In reality, the inelastic dark matter halo will free-fall only for a short instance before it undergoes heating due to shocks. Nevertheless, the details of collapse during this period can be approximated by the free-fall picture since the halo will virialize regardless and the analysis of fragments is agnostic of the specific details during this period.}: Inelastic dark matter overdensities free-fall into the gravitational potential set by mostly CDM. The CDM particles form an NFW-like halo since the particles can pass through the center of the halo without interacting. However, since inelastic dark matter particles are collisional on galactic scales, the inelastic dark matter halo free-falls with $t_\mathrm{collapse} \ll t_\mathrm{cool}$, so that the collapse trajectory is $d \log T_{\chi}/ d \log n_{\chi} \simeq 2/3$. As the inelastic dark matter halo becomes hotter and denser, cooling starts to become efficient.
    \item {\bf Nearly virialized contraction}: The inelastic dark matter halo eventually virializes due to the kinetic pressure of the inelastic dark matter particles as its temperature increases. The trajectory transitions from the free-fall regime into the nearly virialized contraction regime. At this stage of collapse, the trajectory follows the constant Jeans mass contour with $M = m_J$, the Jeans mass in Eq. \eqref{eq:m_J}. Indeed, differentiating Eq. \eqref{eq:m_J} and requiring that $m_J$ is constant yields a slope of the collapse trajectory of $d \log T_\chi / d \log \rho_\chi \simeq 1/3$, which when combined with Eq. \eqref{eq:DE},  requires a collapse time of $t_\mathrm{collapse} \simeq t_\mathrm{cool}/6$.
    \item {\bf Fragmentation}: Due to the quasi-static collapse of the DE halo, the density and the cooling rate both increase. At sufficiently large densities, when $t_\mathrm{cooling} \simeq t_\mathrm{ff}$, cooling is so efficient that the gas collapses at the free-fall time and the trajectory closely follows the contour set by this equality. This trajectory leads to a decrease in the Jeans mass, resulting in fragmentation of the inelastic dark matter halo. Fragmentation continues until one of two stop conditions are reached:
    \begin{enumerate}
        \item The fragments become optically thick, and so further cooling is inefficient. However, these objects may continue to collapse due to surface cooling, which we do not treat in this work.
        \item The fragments become sufficiently dense so that the dominant source of pressure is the dark photon repulsive force. Hence the Jeans mass becomes independent of the temperature and further cooling does not decrease the Jeans mass -- resulting in a pressure supported compact object.
    \end{enumerate}
\end{enumerate}

Therefore the three stages of collapse can be encapsulated in the collapse time for the inelastic dark matter halo as
\begin{align}
  t_\text{collapse} = 
  \setlength{\arraycolsep}{0pt}
  \renewcommand{\arraystretch}{1.2}
  \left\{\begin{array}{l @{\quad} l @{\quad} r l}
        t_\mathrm{ff} = (16 \pi G \rho_{\chi})^{-1/2} & M>m_J, & \text{Adiabatic free-fall} \\
        \frac{1}{6} t_\text{cool} & M = m_J \text{ and } t_\mathrm{cool} > t_\mathrm{ff}, & \text{Nearly virialized contraction} \\
        t_\mathrm{ff} & t_{\mathrm{cool}} \simeq t_\mathrm{ff}. &\text{Fragmentation}
  \end{array}\right.
  \label{eq:t_collapse}
\end{align}

\subsection{Collapse Trajectories}

\begin{figure}[tbp]
\centering 
\includegraphics[width=.9\textwidth]{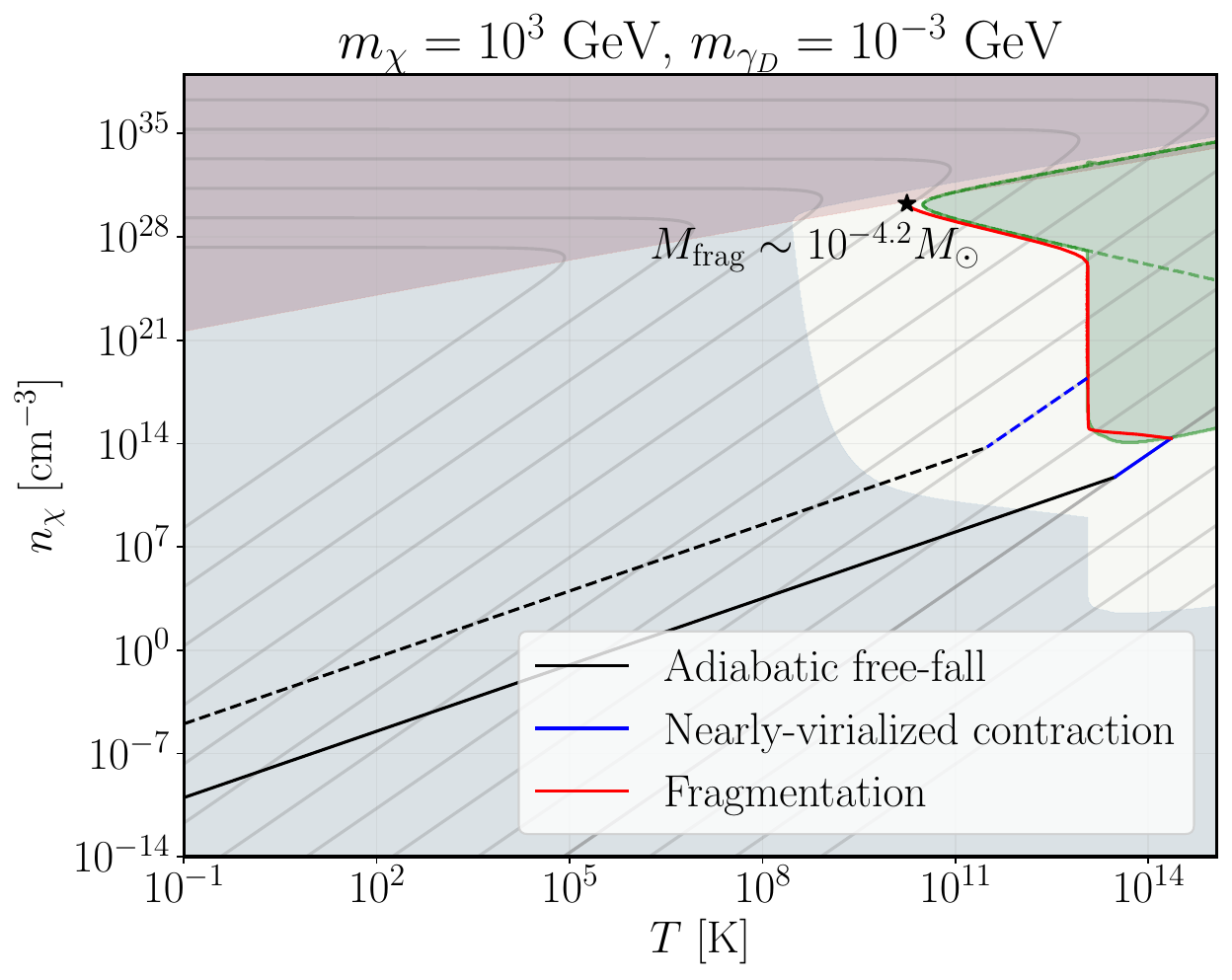}
\caption{Trajectory plot for the evolution of the inelastic dark matter halo. Two different trajectories are shown with different initial conditions: the solid lines correspond to a trajectory with $M = 10^{10} M_\odot$ and the dashed lines correspond to a trajectory with $M = 10^6 M_\odot$. The blue shaded region is where $t_\mathrm{cool} > H_0^{-1}$, the green shaded region is where $t_\mathrm{ff} > t_\mathrm{cool}$ with its boundary representing where $t_\mathrm{ff} = t_\mathrm{cool}$, and the red shaded region is where the halo is optically thick to Bremsstrahlung. The dashed green line in the green contour represents the contour boundary if there was only cooling from Bremsstrahlung. The faint gray curves correspond to constant Jeans mass contours. The final fragment mass is $M_\mathrm{frag} \sim 10^{-4.2} M_\odot$ and is indicated by a black star at the end of the red curve. For these parameters fragmentation ends once the fragments become optically thick to both Bremsstrahlung and inelastically produced dark photons. Details in text.  \label{fig:bs_finish}}
\end{figure}

With the piecewise definition of the collapse time $t_\mathrm{collapse}$ in Eq. \eqref{eq:t_collapse}, the three stages of collapse are completely encapsulated in $t_\mathrm{collapse}$, while the entire process is governed by Eq. \eqref{eq:DE}. This simplified treatment is advantageous because we only have to consider three separate regions for our nonlinear collapse. Furthermore, since the evolution of the inelastic dark matter halo evolves with a constant slope in log-log temperature-density parameter space for adiabatic free-fall and nearly-virialized contraction stages, we need only solve Eq. \eqref{eq:DE} during the fragmentation stage.

Fig. \ref{fig:bs_finish} shows two different possible trajectories for the non-linear collapse of an inelastic dark matter halo. Before discussing the individual trajectories of the inelastic dark matter halo, we first explain the shaded regions. The blue shaded region is where the cooling timescale is longer than the Hubble time. In other words, this is the region where $t_\mathrm{cool} > H_0^{-1}$, where $H_0$ is the Hubble parameter which we took arbitrarily to be $H_0 = 70$ km/s/Mpc. The inelastic dark matter halo trajectory must be outside of this contour by the time it reaches the nearly-virialized contraction phase in order for it to have enough time to cool. The green shaded contour is where cooling is very efficient, so that $t_\mathrm{ff} > t_\mathrm{cool}$. The boundaries of this contour represent where $t_\mathrm{ff} = t_\mathrm{cool}$, while the dashed green lines in the contour would be the boundaries of the contour given only cooling from Bremsstrahlung. The red shaded region represents where the inelastic dark matter is optically thick to Bremsstrahlung dark photons. In this case, the Bremsstrahlung process is no longer efficient and no further cooling can occur. We note that in the parameter space we considered, the optically thick region for dark photons from the inelastic processes were never the stopping condition for our collapse since either the inelastic dark matter halo would cool into a regime where Bremsstrahlung cooling was dominant, or it would become dark photon pressure supported. Finally, the faint gray curves represent contours of constant Jeans masses, each separated by two orders of magnitude. The smaller constant Jeans mass curves are on the left-side of the plot while the larger Jeans mass curves are on the right-hand side.

Moving onto the trajectories themselves, the solid lines in Fig. \ref{fig:bs_finish} correspond the trajectory where the initial mass of the inelastic dark matter halo is $M = 10^{10} M_\odot$. Starting at some initial density and temperature, the halo first free-falls due to its self-gravity (indicated in black) before entering the nearly-virialized contraction phase (indicated in blue) once its Jeans mass increases to the point where $m_J \simeq M$. It then evolves along the constant Jeans mass contour of $m_J = 10^{10} M_\odot$ until the trajectory intersects the $t_\mathrm{ff} \simeq t_\mathrm{cool}$ border where cooling has become so efficient that the halo starts to fragment (indicated in red). The intersection is in the inelastic cooling-dominated region, and so fragmentation initially starts via inelastic cooling. At this point, the trajectory goes through the $t_\mathrm{ff} > t_\mathrm{cool}$ contour instead of following the outline set by the contour. This is due to the fact that the contour is no longer monotonically decreasing in number density as the temperature increases\footnote{This is one major difference between inelastic and Bremsstrahlung cooling. Bremsstrahlung cooling during fragmentation is always monotonically decreasing in number density with increases in temperature; $cf.$ the underlaid dashed curve inside the green $t_\mathrm{ff} > t_\mathrm{cool}$ contour in Figure \ref{fig:bs_finish}.}. However, even if cooling is so efficient that $t_\mathrm{ff} > t_\mathrm{cool}$, the collapse time is limited by the free-fall timescale. Without this constraint, if the trajectory were to follow the $t_\mathrm{ff} = t_\mathrm{cool}$ contour in the inelastic cooling-dominated regime, then the inelastic dark matter halo would be cooling and expanding. The fragmentation phase continues, and Bremsstrahlung cooling eventually becomes the dominant cooling channel. Fragmentation continues under Bremsstrahlung cooling until it reaches one of the end conditions as previously described. In this case, the inelastic dark matter halo becomes opaque to Bremsstrahlung-produced dark photons before reaching the pressure-supported region (indicated by the black star in Fig. \ref{fig:bs_finish}). The final fragment mass is given by $M_\mathrm{frag} \sim 10^{-5.8} M_\odot$ and is indicated by a black star on the figure. Since the stopping condition was due to an optical thickness condition rather than a pressure-supported condition, this fragment is not the final state of the compact object and can continue to cool due to, for instance, surface cooling. Evolution of these fragments after the end of fragmentation would require a more careful analysis.

Meanwhile, the dashed lines in Fig. \ref{fig:bs_finish} correspond to the trajectory for an inelastic dark matter halo with a lower initial mass of $M = 10^6 M_\odot$ and a higher initial number density. The free-fall period proceeds as before, and the inelastic dark matter halo enters the nearly-virialized phase occuring at a much lower temperature due to the fact that it has a lower initial mass (and hence the $m_J = M$ intersection occurs at lower temperatures). It continues in this regime until it reaches the $t_\mathrm{ff} = t_\mathrm{cool}$ border at a point where the border is nearly vertical corresponding to a sharp drop in cooling at the threshold temperature for inelastic transitions, and so the inelastic dark matter halo fragments at a roughly constant temperature $T \sim \delta$. Further up the trajectory, as the densities of the fragments become sufficiently high, Bremsstrahlung cooling becomes the dominant process as in the case of the $10^{10}~M_\odot$ trajectory, and inelastic cooling effectively shuts off. Once again, the halo continues to cool and fragment due to Bremsstrahlung until it stops when it becomes optically thick. As the end parts of the trajectory as well as the ending conditions for the two trajectories are the same, the resulting masses of the final fragments are also the same. This implies then, that the resulting landscape of final fragments would be agnostic to both the initial conditions and the formation histories of the inelastic dark matter halos.

\begin{figure}[tbp]
\centering 
\includegraphics[width=.9\textwidth]{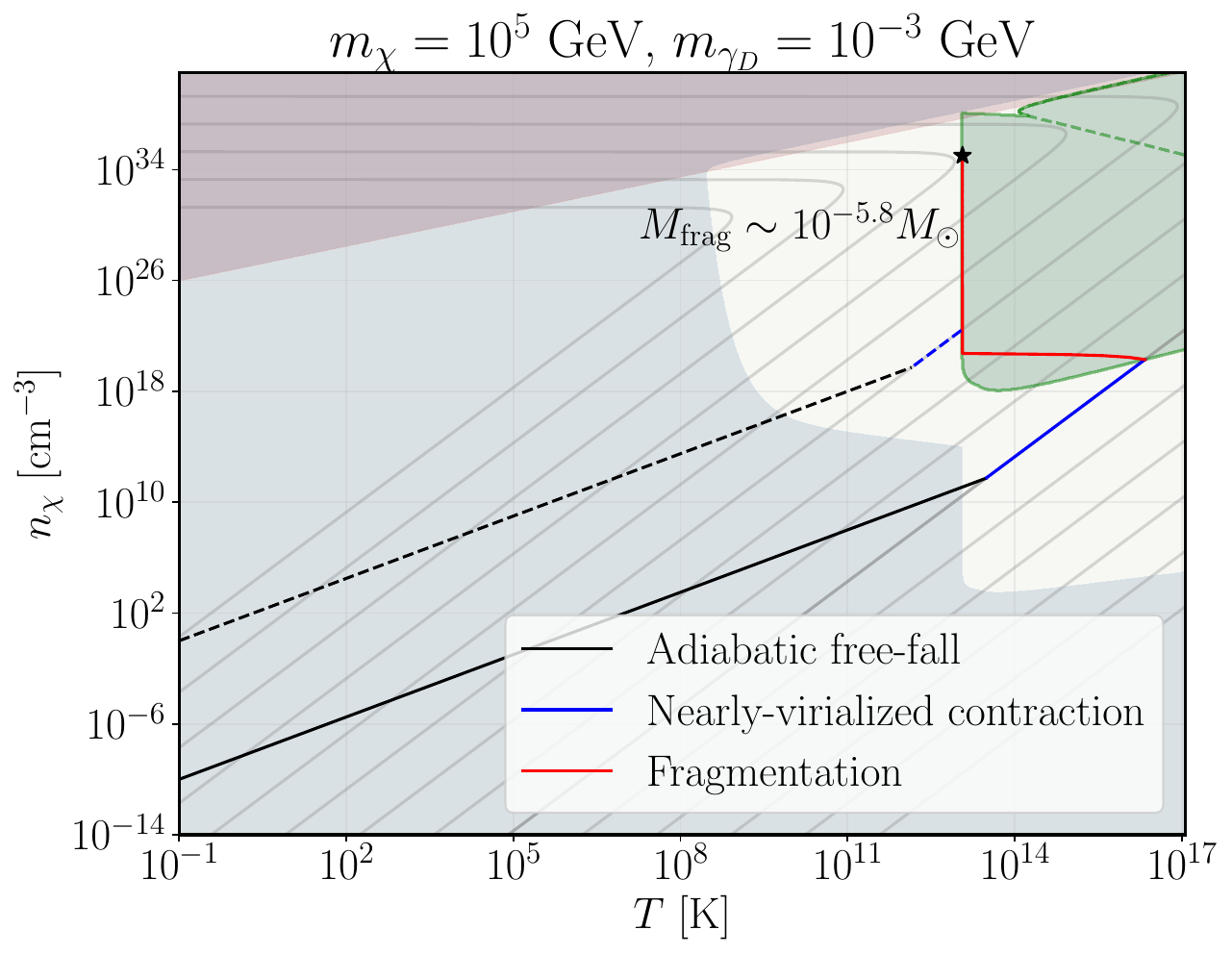}
\caption{Trajectory plot for the evolution of the inelastic dark matter halo. Two different trajectories are shown with different initial conditions: the solid lines correspond to a trajectory with $M = 10^{6} M_\odot$ and the dashed lines correspond to a trajectory with $M = M_\odot$. The blue shaded region is where $t_\mathrm{cool} > H_0^{-1}$, the green shaded region is where $t_\mathrm{ff} > t_\mathrm{cool}$ with its boundary representing where $t_\mathrm{ff} = t_\mathrm{cool}$, and the red shaded region is where the halo is optically thick to Bremsstrahlung. The dashed green line in the green contour represents the contour boundary if there was only cooling from Bremsstrahlung. The faint gray curves correspond to constant Jeans mass contours. The final fragment mass is $M_\mathrm{frag} \sim 10^{-4.2} M_\odot$ and is indicated by a black star at the end of the fragmentation (red) trajectory. Fragmentation ends in this scenario due to it becoming pressure supported. Details in text. \label{fig:ie_finish}}
\end{figure}

While Fig. \ref{fig:bs_finish} showed possible trajectories in the case where the model parameters ($i.e.$ $m_\chi, m_{\gamma_D}, \alpha_D, \delta$) were chosen so that the temperature intervals at which one cooling channel was dominant over the other were separate, one can also have the two temperature regimes overlap. Fig. \ref{fig:ie_finish} instead considers two possible trajectories for a higher inelastic dark matter mass of $m_\chi = 10^5$ GeV. In this case, as evident from the shape of the $t_\mathrm{ff} = t_\mathrm{cool}$ contour, the inelastic cooling channel is always the dominant cooling channel and Bremsstrahlung cooling can effectively be ignored. In this scenario, the inelastic dark matter halo collapses and cools as if the Bremsstrahlung channel was not present. Similar to before, the solid lines correspond to a higher initial mass of $M = 10^{6} M_\odot$ while the dashed lines correspond to a trajectory with a lower initial inelastic dark matter halo mass of $M = M_\odot$. The fragmentation process for the solid set of curves again happen through the $t_\mathrm{ff} > t_\mathrm{cool}$ contour (this time, for a larger portion of fragmentation compared to Fig. \ref{fig:bs_finish} since the halo starts fragmenting at a much higher temperature) due to the fact that the collapse of the inelastic dark matter halo is limited by the free-fall time, although cooling is extremely efficient. Fragmentation continues in the inelastic cooling-dominated region until it stops due to the fragments becoming pressure-supported (indicated by the black star in Fig. \ref{fig:ie_finish}). Regardless of the starting initial conditions, the masses of the final fragments are again the same as long as the trajectory is able to reach the $t_\mathrm{ff} = t_\mathrm{cool}$ contour.

Before ending this section, we discuss the initial conditions for the trajectories in both Figs. \ref{fig:bs_finish} and \ref{fig:ie_finish}.  Realistically, the initial conditions of the inelastic dark matter halo in the non-linear collapse phase would be set by the size of the perturbation, the fraction of all of dark matter that lies in these inelastic dark matter particles, as well as the starting temperature of the gas. However, we see that the resulting analysis of the final dark matter gas fragments seem to be relatively agnostic to the initial conditions. In addition, we note that as a rough estimate, the solid lines in both figures approximately correspond to initial conditions where the inelastic dark matter particles were homogeneously distributed in the entire overarching Milky Way-sized halo at turnaround with radius $\lambda_\mathrm{MW}/2 \approx 2 $ Mpc with $f \simeq 0.01$ fraction of the dark matter as the inelastic dark matter we have studied here, assuming an arbitrary starting temperature of $T=0.1$ K. Meanwhile, the dashed lines, which correspond to the smaller initial masses, were required to start with higher number densities ($i.e.$ by taking a large perturbation length) in order to transition into the nearly-virialized contraction phase outside of the inefficient cooling region. These trajectories demonstrate scenarios where the inelastic dark matter halo is just hot enough that it cools via inelastic cooling so that fragmentation proceeds at a nearly constant temperature. In future work, it may be interesting to explore cosmologies with a more varied set of initial dissipative dark matter distributions, including possible effects of initial clustering within the galactic halo. The treatment we have presented here could still be applied to a richer comsological scenario with more varied initial dark matter distributions, since this treatment only requires an initial dark matter mass, density, and temperature as starting inputs.

\section{Distribution of Dark Compact Objects} \label{sec:landscape}

\subsection{Final Fragmentation Masses}

\begin{figure}[tbp]
\centering 
\includegraphics[width=\textwidth]{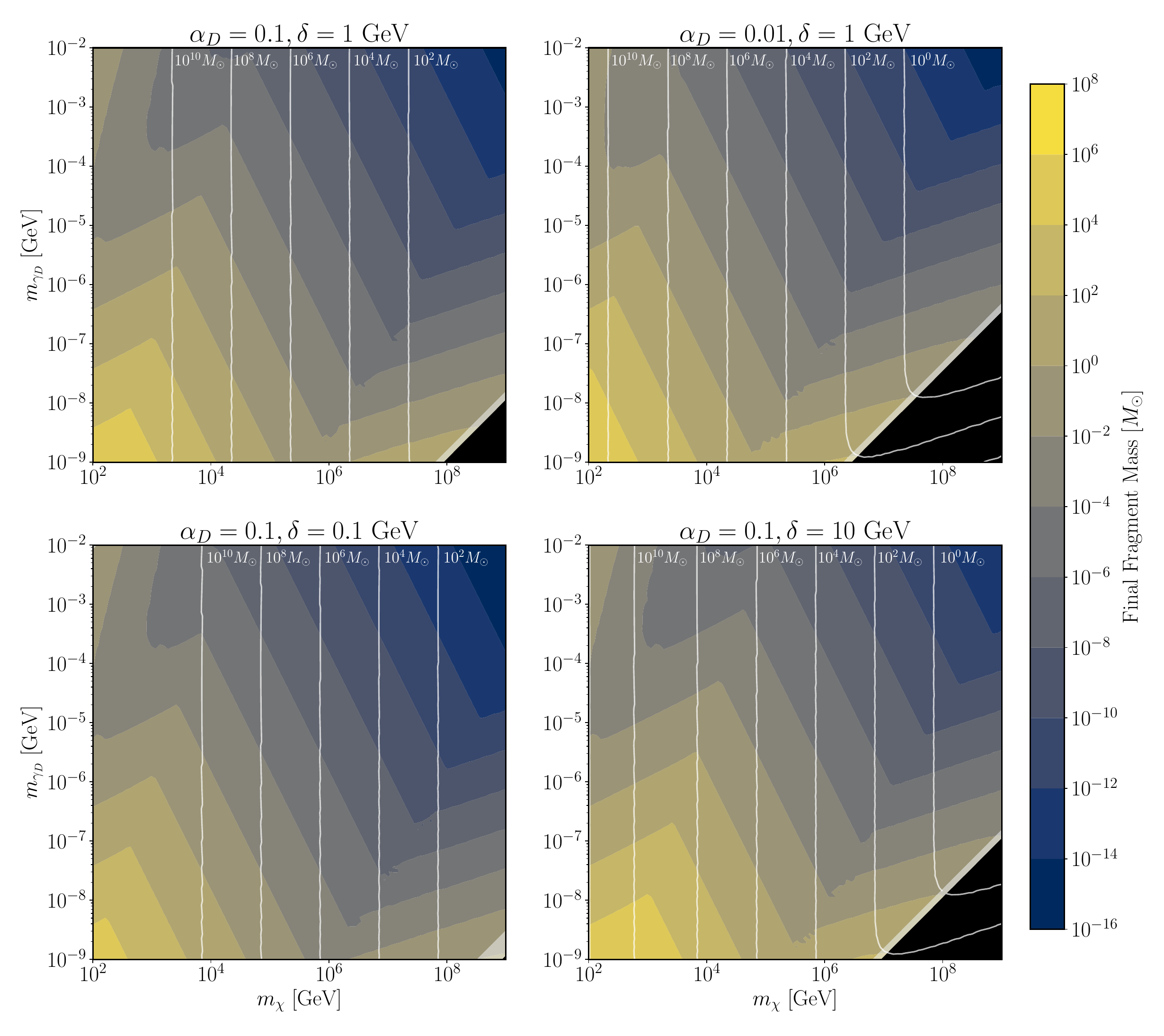}
\caption{Final fragmentation mass contours in terms of the inelastic dark matter mass $m_\chi$ and the dark photon mass $m_{\gamma_D}$. In each of the panels, choices of the dark coupling $\alpha_D$ and the mass splitting $\delta$ are indicated. Brighter regions correspond to higher fragment masses and darker regions correspond to lower masses. The gray shaded region near the bottom right of the panels represents locations where the stabilized fragments are black holes following Eq. \eqref{eq:C_max}. Black shaded regions seen in the lower right corners of each panel indicate regions where the inelastic dark matter halo begins to fragment only after it reaches a temperature of $T_\chi^\mathrm{max} = m_\chi/10$, meaning the inelastic dark matter particles would be semi-relativistic. We expect black holes to also form in this black parameter space, which is comparable to the ``runaway'' black hole parameter space studied in \cite{Chang:2018bgx}, but strictly speaking this requires further analysis, since much of this parameter space involves relativistic dynamics, and we have assumed non-relativistic dark matter in our cooling equations. The white solid lines indicate the maximum initial mass of dark matter halos that would fragment in a non-relativistic regime along its cooling trajectory. More details can be found in the text.
\label{fig:m_ie}}
\end{figure}

In the previous section we sketched out some possible individual trajectories for an inelastic dark matter halo given various initial conditions. However, it is important to note that as long as the inelastic dark matter halo reached the third phase of non-linear collapse ($i.e.$ fragmentation), the resulting final fragment was independent of initial conditions such as the temperature or density of the halo. Therefore, assuming that the halo began to fragment, one can acquire a landscape of final fragmentation masses that is agnostic of the initial conditions, and instead is only dependent on the parameters in the particle physics model (which in our case, are $\alpha_D, \delta, m_\chi$ and $m_{\gamma_D}$).

Fig. \ref{fig:m_ie} shows several possible distributions of the final fragmentation masses for a various selection of model parameters. In all of the panels, the range for the inelastic dark matter mass was between $10^2$ GeV and $10^9$ GeV, while the dark photon masses were between $10^{-9}$ GeV and $ 10^{-2} $ GeV. Since we chose to work in the limit where $m_{\gamma_D} \ll \delta $, the values for the mass splitting $\delta$ were taken to be $\delta = 0.1, 1$ or $10$ GeV. We considered coupling strengths of $\alpha_D = 0.1$ and $\alpha_D = 0.01$. Each panel in Fig. \ref{fig:m_ie} are snapshots taken at the specific value of $\alpha_D$ and $\delta$, indicated by the plot titles. The final fragment masses were taken at the end of fragmentation as described in the previous section, which correspond to the black stars plotted in Figs. \ref{fig:bs_finish} and \ref{fig:ie_finish} at the end of the fragmentation trajectory. Due to the fact that we had a non-relativistic expression for the Bremsstrahlung and inelastic scattering cross sections (and hence the cooling rates), we limited the maximum possible temperature of the inelastic dark matter gas to be $T_\chi^\mathrm{max} = m_\chi/10$ so that the gas remains non-relativistic. By intersecting the $t_\mathrm{cool} = t_\mathrm{ff}$ contour with the maximum allowed temperature, one can find the associated maximum initial mass of the inelastic dark matter halo which allows for fragmentation. These initial maximum masses are indicated as solid white lines with their corresponding labels.

Furthermore, the shaded gray regions in Fig. \ref{fig:m_ie} represent regions where the final fragments are black holes. One can estimate whether not these fragments would be black holes or pressure-supported dark stars\footnote{Note that what we refer to as dark stars are fragments that are dark pressure-supported. No nuclear physics is involved.}. To estimate whether the final fragments would be sufficiently dense and compact enough to form a black hole, one can compute the compactness $C$ of the object,
\begin{align}
    C = \frac{GM}{R},
\end{align}
where $M$ is the final mass of the fragment and $R$ is its size, given by $R \sim \lambda_J$ since fragmentation occurs at the objects' Jeans scales. A compactness of $C = 1/2$ is the maximum value of the compactness parameter, which corresponds to a black hole, since for this value $R$ is the Schwarzschild radius. For compact objects that stop fragmenting due to being supported by the dark photon pressure, one can compute the compactness at this point since further cooling will not cause the object to collapse further. However, since the fragments which stop fragmenting due to becoming optically thick can still continue to collapse due to surface cooling, one can acquire an estimate for the maximal compactness of the fragments by assuming that the exotic compact objects will continue to evolve to the point where they are supported by the dark photon repulsive force. The compactness of the fragments at this point can be estimated by
\begin{align}
    C = \frac{m_\chi m_{\gamma_D} M}{\pi} \left( \frac{G^3}{\alpha_D} \right)^{1/2}, \label{eq:C_max}
\end{align}
which can be acquired by taking the limit where $m_J \simeq M$ and neglecting the subdominant kinetic pressure term in $m_J$ \cite{Chang:2018bgx}.

Meanwhile, the black shaded regions indicate where -- without a relativistic description of the cooling processes -- the initial gas of inelastic dark matter particles would not be able to fragment at all because the gas becomes pressure supported due to the dark photon radiation pressure before fragmentation is able to begin. In other words, the trajectory of the inelastic dark matter halo never reaches the green $t_\mathrm{ff} = t_\mathrm{cool}$ contours in Figs. \ref{fig:bs_finish} and \ref{fig:ie_finish} for any given initial inelastic dark matter gas mass. While we expect more fragments to end up as black holes inside of the shaded black regions, we are unable to say, with certainty, that these fragments would evolve in a similar manner in the relativistic regime as they do in the non-relativistic regime. We leave the relativistic treatment of our cooling mechanisms for a future study.

However, we note that the additional cooling channel opens up new parameter space for non-relativistic final fragments when compared to a scenario with only Bremsstrahlung cooling. To see this, one can infer from Fig. \ref{fig:bs_finish} that in order to intersect the Bremsstrahlung $t_\mathrm{ff} = t_\mathrm{cool}$ contour at the same temperature, one would require much higher number densities for cooling to become efficient. These number densities may never have been reached if we stayed in the non-relativistic regime with only Bremsstrahlung. Therefore the addition of inelastic cooling allows some of the inelastic dark matter halos to fragment where they could not have before.

The shape of the final fragment mass distribution in Fig. \ref{fig:m_ie} is ultimately determined by the shape of the fragmentation contours ($i.e.$ the borders of the green contours) in Figs. \ref{fig:bs_finish} and \ref{fig:ie_finish}. The behaviour of the final fragment masses towards the left hand-side of the panels corresponding to lower inelastic dark matter masses in Fig. \ref{fig:m_ie} are all due to a fragmentation trajectory that ends with Bremsstrahlung cooling ($i.e.$ a scenario such as Fig. \ref{fig:bs_finish}). Meanwhile, everything to the right corresponds to scenarios where the inelastic channel is the dominant process throughout fragmentation, and so fragmentation ends due to the object becoming pressure supported. The negatively sloped region is where fragmentation ends while the inelastic dark matter halo temperature is near the activation energy of the inelastic cooling channel $T\sim \delta$ ($i.e.$ a scenario like Fig. \ref{fig:ie_finish}). In this region, as soon as the fragments heat up due to collapse, it cools immediately through inelastic processes to keep the fragment at a roughly constant temperature. The region to the right where the constant fragment contours are positively sloped is where the inelastic dark matter halo becomes pressure-supported when $t_\mathrm{ff} > t_\mathrm{cool}$, which occurs inside of the green contour in Fig. \ref{fig:ie_finish}. Therefore, due to the inelastic cooling channel being several orders of magnitude more effective than the Bremsstrahlung channel,  we can conclude that the additional cooling mechanism allows for much smaller final fragment masses when compared to the scenario with only Bremsstrahlung cooling as in Ref. \cite{Chang:2018bgx}. 

All of the presented fragment masses in Fig. \ref{fig:m_ie} serve as an estimate for the lower bound of the dark compact objects since they may continue to collapse for various reasons -- for example, by surface cooling or being involved in merger events. Our treatment is valid to study these fragments (which are analogous to protostars in the baryonic sector) as isolated systems, and more careful simulations must be performed to properly study the evolution of these dark stars past the end of fragmentation.

\subsection{Observational Prospects}

Now we briefly discuss avenues to observe these dark exotic compact objects. In this study, we have not enforced any non-gravitational interactions between the baryonic sector and our subdominant dark sector. Hence we will focus on the non-gravitational signatures for the compact objects. The exotic compact objects in this work can be studied and searched for in a similar manner to massive compact halo objects (MACHOs) and primordial black holes through microlensing experiments \cite{EROS-2:2006ryy, OGLEIV, Katz:2019qug, Croon:2020wpr}. In addition, Pulsar Timing Arrays (PTAs) could detect exotic compact objects through single transit event searches \cite{Dror:2019twh} as well as the effect of the compact object distribution on the small-scale power spectrum \cite{Lee:2020wfn}.

With the massive amounts of resources being allocated to gravitational wave astronomy, gravitational waves are exciting probes which may also provide insight to search for dark compact objects. In particular, as mentioned in the introduction, the existence of sub-solar black holes would have significant implications for astrophysics and cosmology due to the lack of a conventional production mechanism for low mass black holes \cite{Shandera:2018xkn}. While ongoing searches for sub-solar mass black holes have not yielded any detections, the null detection of signals from sub-solar black holes can be used to put constraints on the model parameters of a dissipative dark sector \cite{LIGOScientific:2022hai}. In the future, both ground-based experiments such as the Einstein Telescope \cite{Maggiore:2019uih} and space-based experiments such as the Laser Interferometer Space Antenna (LISA) \cite{LISA} are expected to be able to detect these exotic compact objects \cite{Kuhnel:2018mlr, Shandera:2018xkn, Diamond:2021dth, LISA:2022kgy}. However, to study the gravitational wave signatures of these exotic compact object binaries, more detailed calculations must be done to analyze the behaviour of the fragments after the end of fragmentation.

We end this section by commenting on the fact that we have ignored any constraints on the model parameters arising from cosmological sources. The goal of this work was not to study the specific particle physics model and constrain the associated model parameters, but rather to investigate the phenomenological implications of having multiple cooling channels in the formation history of exotic compact objects. Therefore, these cosmological constraints, such as ones from the early universe, as considered in Ref. \cite{Chang:2018bgx} for the simple dark electron and dark photon model, and Ref. \cite{Bansal:2022qbi} for atomic dark matter, are not considered in this study. While one can perform similar analyses to these studies for our specific model, we leave this exploration for future work. 

\section{Conclusion} \label{sec:conclusion}

In this work we studied the effect of having multiple cooling channels on the evolution of dark exotic compact objects without the use of complicated and expensive numerical simulations. By considering a subdominant, dissipative dark sector comprised of a dark Dirac fermion and a dark photon, where the fermion undergoes a mass splitting, we showed that a gas of inelastic dark matter particles and dark photons can collapse under the influence of several cooling channels. The addition of multiple cooling mechanisms were shown to drastically modify both the formation history and the distribution of dark exotic compact objects, such as dark stars and dark black holes, when compared to the scenario with only one relevant cooling mechanism.

This work is a follow-up to Ref. \cite{Chang:2018bgx}, which studied the formation and evolution of dark exotic compact objects under one relevant cooling channel, dark Bremsstrahlung. In the regions where Bremsstrahlung was the dominant cooling channel, we were able to reproduce their results for both the individual trajectories of dissipative dark halos and the resulting distribution of final fragment masses. However, in the regions where the new inelastic cooling channels were dominant over Bremsstrahlung cooling, we obtained significantly different evolution histories for the dark halos and their resulting distributions of final fragment masses. Namely, since the inelastic cooling rates were much larger in magnitude than the Bremsstrahlung cooling rates, cooling becomes efficient at much lower number densities if the temperature of the inelastic dark matter halo was in the right regime. For model parameters where the inelastic cooling channel was dominant throughout the collapse, we were able to obtain much lower fragment masses than what would have been obtained with just Bremsstrahlung. Hence having multiple cooling channels results in two significant differences: due to the enhanced cooling, our non-relativistic treatment was valid for more of the parameter space when compared to the scenario with one cooling channel, and we found the inelastic model resulted in much more parameter space with smaller mass fragments, including sub-solar mass black holes. 

While our analysis was able to acquire a simplified estimate for the possible distribution of dark exotic compact objects in a subdominant, dissipative dark sector with multiple cooling channels, more extensive numerical studies and modelling are required to understand these structures further. Additionally, our treatment for the inelastic dark matter gas was non-relativistic. As we saw in the resulting fragment distributions, to fully model all of the parameter space, we require a relativistic treatment of the cooling interactions. We leave these interesting ideas for a future investigation.

\acknowledgments

We thank Daniel Egana-Ugrinovic, Sarah Shandera, and Aaron Vincent for useful discussions. JLK is supported by the Natural Sciences and Engineering Research Council of Canada (NSERC) through a CGS-D. The computations in this work were performed on the computing clusters at the Centre for Advanced Computing (CAC) at Queen's University, supported in part by the Canada Foundation for Innovation. This work was supported by NSERC and the Arthur B. McDonald Canadian Astroparticle Physics Research Institute. This research was also supported in part by Perimeter Institute for Theoretical Physics. Research at Perimeter Institute is supported by the Government of Canada through the Department of Innovation, Science and Economic Development Canada and by the Province of Ontario through the Ministry of Colleges and Universities.

\bibliographystyle{JHEP}
\bibliography{refs}

\end{document}